# Thermodynamic origin of the magnetization peak effect in the superconductor $Nb_3Sn$

R. Lortz[1], N. Musolino[1], Y. Wang[1], A. Junod[1], N. Toyota[2]

[1]Department of Condensed Matter Physics, University of Geneva, 24 Quai Ernest-Ansermet, CH-1211 Geneva 4, Switzerland

[2]Physics Department, Graduate School of Science, Tohoku University, 980-8571 Sendai, Japan

We report a pronounced peak effect in the magnetization and the magnetocaloric coefficient in a single crystal of the superconductor $Nb_3Sn$. As the origin of the magnetization peak effect in classical type-II superconductors is still strongly debated, we performed an investigation of its underlying thermodynamics. Calorimetric experiments performed during field sweeps at constant temperatures reveal that the sharp increase in the current density occurs concurrently with additional degrees of freedom in the specific heat due to thermal fluctuations and a liquid vortex phase. No latent heat due to a direct first-order melting of a Bragg glass phase into the liquid phase is found which we take as evidence for an intermediate glass phase with enhanced flux pinning. The Bragg glass phase can however be restored by a small AC field. In this case a first-order vortex melting transition with a clear hysteresis is found. In the absence of an AC field the intermediate glass phase is located within the field range of this hysteresis. This indicates that the peak effect is associated with the metastability of an underlying first-order vortex melting transition.

## I. INTRODUCTION

In some 'classical' low-$T_c$ type-II superconductors such as $CeRu_2$ [1,2,3], $NbSe_2$ [1], $V_3Si$ [4], $MgB_2$ [5] and certain heavy-fermion superconductors including $UPt_3$ [6] and $UPd_2Al_3$ [7], a sharp peak effect in the magnetization related to an abrupt change from reversible to irreversible behaviour is reported along their upper critical field line [8,9]. The most important question regarding this enhanced flux pinning is whether its origin is a dynamical change in the vortex rigidity or a phase transition in the thermodynamic sense. Theoretical and experimental studies suggest the existence of two solid phases of vortex matter; For example the peak effect has been described as a transformation of a quasi-ordered Bragg glass into a highly disordered phase [9]. Although various models have been suggested, there is currently no consensus regarding the underlying nature of the disordered phase and the corresponding order-disorder transition.

Recently, we discovered a signature of thermal fluctuations and vortex melting in the specific heat of a high purity $Nb_3Sn$ single crystal. The sample furthermore showed a pronounced peak effect in the magnetization, which motivated us to study whether there is a link between the peak effect and the onset of thermal fluctuations.

## II. EXPERIMENTAL TECHNIQUES

To do so, we compare the results of three complementary thermodynamic quantities, the magnetization $M$, the isothermal magnetocaloric coefficient $M_T$ and the specific heat $C$ which are increasingly less sensitive to irreversibility, as we will demonstrate respectively below. The calorimetric experiments were performed by a quasi-isothermal heat-flow calorimeter [10,11]. In this method, the sample is linked to a heat sink by means of a sensitive heat-flow meter of high thermal conductance k. This is achieved by using a sample platform made of sapphire suspended on a thermopile of 24 Au-Fe/Chromel thermocouples. To measure the specific heat under similar conditions to the isothermal magnetization a thin-film heater is deposited on the sample platform. An AC heat-flow technique can then be used during field sweeps at constant temperatures with a small temperature modulation of the order of ~1 mK [12]. The specific heat is a pure thermodynamic bulk quantity, insensitive to irreversible effects due to flux pinning and hence a perfect tool for revealing the presence of phase transitions in the thermodynamic sense. The design of the calorimeter also allowed us to measure a second quantity, the isothermal magnetocaloric coefficient, by applying a DC heat-flow technique during a field sweep. The sample we used is a high quality $Nb_3Sn$ single crystal with a $T_c$ width of ~20 mK (dimensions 3 x 1.3 x 0.4 $mm^3$) [13]. The magnetization was measured with a commercial vibrating sample magnetometer (VSM) during field sweeps of 0.25 T/min with vibration frequency 82 Hz.

## 3. EXPERIMENTS
### 3.1. ISOTHERMAL MAGNETIZATION

Fig. 1a shows magnetization measurements at various fixed temperatures below $T_c$. The units J gat$^{-1}$ T$^{-1}$ were chosen for comparison with the calorimetric measurements (1 gat=1/4 mole in Nb$_3$Sn). Apart from a hysteresis separating the ascending and descending branches at low fields, the curves are reversible and the data measured for + and – $dB_a/dt$ merge ($B_a = \mu_0 H$ is the applied field). The upper critical field $H_{c2}$ is marked as a kink above which $M(H)$ reaches a constant value. For $T < 15$ K a small hysteresis loop appears below $H_{c2}$ due to a pronounced peak effect, also observed in the AC susceptibility curve for the same sample [14]. In the presence of some flux pinning, the experimental curves can be understood as a superposition of a reversible thermodynamic contribution which is independent of the direction of the field sweep and an irreversible contribution which is directly linked to the critical current density $J_C$. The sign of the second component depends on the direction of the field sweep, thus resulting in a hysteresis loop between the ascending and descending branches of the magnetization. In Fig. 1b we have separated these two contributions by calculating the thermodynamic component: $M_{rev}= (M^+ + M^-)/2$ and the irreversible component: $M_{irr}=(M^+ - M^-)/2$, where $M^+$ ($M^-$) is the branch of the magnetization of the ascending (descending) branch for $dB_a/dt>0$ ($dB_a/dt<0$), where $B_a$ is the applied field which is swept at a constant rate. The field value at which the critical current density reaches its maximum is marked as $H_p$. The calculation of $M_{rev}$ works only approximatively in the region of the peak effect, revealing some asymmetry in the distribution of critical currents between $M^+$ and $M^-$: this is most probably due to enhanced surface pinning in $M^-$.

### 3.2. ISOTHERMAL MAGNETOCALORIC COEFFICIENT

Fig. 2a shows measurements of the isothermal magnetocaloric effect $M_T(H) = (dQ/dB_a)_{T=const}$. To determine this quantity the field is swept at a constant $dB_a/dt$ and the heat released or absorbed by the sample is measured. The interpretation of this quantity is more straightforward than that of the adiabatic magnetocaloric coefficient $M_S=(dT/dB_a)_S$, since $M_T$ is measured at constant temperature $T$ rather than constant entropy $S$. $M_T$ is closely related to the magnetization $M$; both quantities share the same units (J/gatT = Am$^2$/gat when expressed per gram-atom, A/m when expressed per volume unit). The magnetization contains a reversible thermodynamic component $M_{T\_rev}$ and an irreversible component $M_{T\_irr}$; it can be shown using thermodynamic relations that $M_{T\_rev} = -T(dM/dT)_{Ba}$. For $T > 15$ K the data in ascending and descending field look rather similar. A small peak is found in both curves, superimposed on the jump at $H_{c2}$ which does not depend on the sign of $dB_a/dt$. We thus conclude that the main contribution arises from $M_{T\_rev}$. As can clearly be seen in Fig. 2, the irreversible component $M_{T\_irr}$ enters the magnetocaloric coefficient in the presence of flux pinning and a peak effect very similar to that observed in $M$ is visible for $T<15$ K. The similarity of the irreversible components in both quantities has been discussed in Ref. [15]: While the irreversibility in $M$ arises from different vortex density distributions across the sample upon increasing and decreasing the field, the origin of the irreversibility in $M_T$ lies in the friction of vortices against pinning centres. Friction can only result in a heat release d$Q$>0, irrespective of the direction of the field sweep, thus creating a hysteresis loop between the ascending and descending branches of $M_T$. The similarity of the hysteresis in $M$ and $M_T$ originates from the fact that they are both governed by $J_C$. In Fig. 2b we separated the two components using the same method as applied to the magnetization data. The irreversible component looks very similar to $M_{irr}$; however, in contrast to the magnetization the $M_{T\_irr}$ curves below the peak regime are not fully reversible. An explanation for this difference may be found in the different conditions during the measurements: $M_T$ is measured in DC mode with a fixed sample position whereas in the VSM technique the sample is vibrating in a magnetic field. Even in the presence of only a tiny field gradient the sample is subjected to a small AC field. It has been shown that small pinning barriers can be overcome by superimposing an AC field on the DC field [16,17]. We repeated $M_T$ measurements at a fixed temperature $T$=12 K for various sweep rates $dB_a/dt$ and found that the behaviour in the region of the peak effect is independent of the sweep rate. However, at lower fields the behaviour changes dramatically if the sweep rate is varied. At fast sweep rates (1.4 T/min) $M_{T\_irr}$ always has a positive signature. Heat is thus released over the whole field range which we attribute to the friction of vortices in the presence of weak effective pinning. Weak effective pinning means that sufficient energy is supplied by the strong electrical field $E \sim dB_a/dt$ to overcome the pinning barriers. Using a slower sweep rate (here 0.7 T/min) $M_{T\_irr}$ changes sign at ~7 T and becomes negative in smaller fields. This means that heat is absorbed from the thermal bath within this field range. It seems that the electrical field no longer supplies sufficient energy to overcome the pinning barriers and so the vortex system takes the energy in the form of heat from its surroundings. The irreversible contribution to the magnetocaloric coefficient is thus an ideal tool for investigating the behaviour of vortex matter in the flux creep regime. The observed behaviour shows that flux creep is enhanced in the phase below the peak effect regime, but this enhancement is clearly less important at higher fields within the peak effect region. This indicates that the pinning mechanism changes dramatically in a small regime below $H_{c2}$.

### 3.3. 'QUASI-ISOTHERMAL' SPECIFIC HEAT

Fig. 3. shows the quasi-isothermal specific heat measurements as a function of the magnetic field. $C(H)$ has been measured in the same calorimeter under similar conditions to those used for measuring $M_T(H)$, albeit with a different technique. Here we present $C$ rather than $C/T$ (which is more commonly used in superconductors) as our purpose is to compare the three quantities by using comparable units and not to perform a detailed specific heat analysis of $Nb_3Sn$. The superconducting parameters such as the Sommerfeld constant $g_S$ and the superconducting condensation energy are however in good agreement with previous specific-heat studies on $Nb_3Sn$, e.g. Ref. [18]. To obtain this quantity the sample temperature is modulated periodically [12] with a small amplitude of ~1 mK. Above a certain cut-off frequency both the phase shift and amplitude of this modulation with respect to the excitation power are related to the specific heat. Although we measure this quantity during a field sweep, it originates from the response of the sample to a small temperature change and not primarily due to the field change as is the case for $M_T$. As the curves are measured at constant temperatures the phonon background enters the measurements as a constant factor which is responsible for the shift of the curves with respect to each other. The superconducting contribution $C^{SC}$ can easily be separated by subtracting the constant normal-state specific-heat value which is found above the upper critical field $H_{c2}$ (Fig. 3 inset, data with positive signature). The specific heat curves are fully reversible and do not depend on the sign of $dB_a/dt$. The transition to the normal state is indicated by a jump which becomes increasingly broad in higher fields. Just below the jump a small upturn in the specific heat indicates that additional degrees of freedom due to thermal fluctuations or a vortex liquid phase suddenly appear. They form a small lambda anomaly at a field $H_l$ superimposed on the specific-heat jump at $H_{c2}$. This anomaly becomes increasingly broadened in higher fields concurrently with the $H_{c2}$ jump. As $C = T\, dS/dT$ and $M_{T\_rev} = -T\, dM/dT = T\, dS/dH$, the reversible part of $M_T$ is closely related to $C$ (see Fig. 3 inset for a comparison of $M_{T\,rev}$ and $C^{SC}$) and we conclude that the lambda anomaly has the same origin as the small peak in $M_T$ for $T > 15$ K. We have already investigated this anomaly in detail in a previous article [19] where we studied the specific heat conventionally in constant fields during a temperature sweep. In that article we showed from a comparison with theoretical calculations that the additional degrees of freedom can be ascribed to a vortex melting. The absence of a latent heat indicates that the solid phase is of a glassy nature and the transition continuous. In the present paper we would like to compare this continuous vortex melting transition with the irreversible anomalies in $M$ and $M_T$ to investigate how this transition is related to the peak effect.

## 3.4. COMPARISON OF THE THREE QUANTITIES

For this purpose we show in Fig. 4 the three quantities $M(H)$, $-M_T(H)$ (inverted for clarity) and $C(H)$ at $T = 13.5$ K together in a single plot. All data have been measured under similar conditions at exactly the same temperature. This plot therefore allows us to compare the purely thermodynamic anomalies in $C$ with the irreversible anomalies in $M_T$ and $M$. The peak effect in $M$ and $M_T$ has its maximum $H_p$ exactly at the onset of the small upturn in $C$ which forms the small lambda anomaly at $H_l$, indicating the smooth transition into a liquid vortex phase. The irreversible region extends up to $H_l$ where the maximum of the fluctuation anomaly occurs in $C$. The combination of these three quantities in the regime of the peak effect shows that its origin is most likely due to flux line lattice softening upon approaching the vortex melting transition, driven by the increasing strength of thermal fluctuations. The magnetization is fully reversible in fields below the peak effect regime indicating weak flux pinning and enhanced flux creep, so the model of a dislocation-free Bragg glass phase as proposed in Ref. [9] accurately describes the relatively ordered vortex phase at lower fields. As no first-order transition is found at the onset of the fluctuations below $H_{c2}$, a direct transition from a Bragg glass to a liquid phase can be excluded from our data. It thus seems likely that an intermediate glassy phase subject to enhanced pinning separates the Bragg glass from a liquid phase, as proposed by theory [9].

## 3.5. COMPARISON WITH 'VORTEX SHAKING' EXPERIMENTS

In Ref. [19] we showed that it is possible to restore the Bragg glass in the peak effect regime in temperature sweep experiments by applying a small AC field parallel to the DC field which helps the vortices to reach equilibrium [16,17]. In this case a small spike due to a latent heat was found indicating that the Bragg glass melts directly via a first-order transition into a liquid. In Figure 5 the magnetocaloric coefficient at $T=13.5$ K measured with an AC field of 10 G at a frequency of 1 kHz superimposed on the DC field is shown in comparison with the original curve without AC field. The behaviour close to $H_{c2}$ changes dramatically. The small hysteresis loop disappears and in its place a large spike is found [20]. The anomaly has the same shape as observed in $M_T$ at the first-order vortex melting transition in ReBa$_2$Cu$_3$O$_x$ (Re=Nd, Y) compounds [15,21]. Contrary to the irreversible loops at the peak effect, the sign of the spike does not depend on the sign of $dB_a/dt$, confirming that it originates from the reversible thermodynamic contribution $M_{T\_rev}$. It is therefore most likely to be a signature of the latent heat from a first-order vortex melting transition as already discussed in Ref. 19. From the integration of the area below the peak we obtain a value of D$S$=0.3 ± 0.1 k$_B$/vortex/x (see also Ref. 19), which can be directly compared [22] to that of the layered high-temperature superconductor YBa$_2$Cu$_3$O$_7$ where a value of ~0.4 k$_B$/vortex/layer ≈ 0.4 k$_B$/vortex/x$_c$ has been reported [23]. We thus find a reasonable value for the latent heat of a vortex-melting transition. Interestingly, a first-order hysteresis is found in the vortex melting field (marked as $H_{m\_up}$ and $H_{m\_down}$ dependent on the sign of the sweep rate $dB_a/dt$) and the shape of the spike appears somewhat deformed for $dB_a/dt<0$. The region where the peak effect maximum occurs is located within the field range of this hysteresis. This indicates that the peak effect is associated with the metastability of an underlying first-order vortex melting transition in the vortex matter. As an overview of the various characteristic fields obtained from all the measurements we present a phase diagram in Fig. 6.

# 4. DISCUSSION AND CONCLUSIONS

From our experiments we conclude that the origin of the peak effect is due to the increasing strength of thermal fluctuations upon approaching the $H_{c2}$ line. This 'thermal disorder' is responsible for a sudden loss of shear modulus rigidity, as already proposed by Giamarchi et al. in a scenario of a sudden proliferation of dislocations close to the vortex melting transition [9]. The vortex solid thus becomes unstable close to the $H_{c2}$ line resulting in a continuous vortex melting transition. In the temperature range where this crossover from a solid to a liquid phase takes place, pinning is strongly enhanced and the magnetization peak effect is found in the form of a sharp increase in the critical current density. In the absence of disorder - or if the disorder is rendered ineffective by a small AC field ('vortex shaking') – our experiments show that thermal fluctuations melt the Bragg glass via a first-order transition into the liquid phase.

In the presence of disorder the Bragg glass thus transforms into a strongly pinned and hence highly disordered intermediate glassy phase, separating the Bragg glass from the liquid. This glassy phase is bound to the region within the hysteresis of the underlying first-order melting transition. We stress that no sharp anomalies due to phase transitions are observed in the specific heat in this case. The irreversible regime extends over the smooth crossover from a solid to liquid phase indicated by the upturn in the specific heat just before the jump at $H_{c2}$. The absence of sharp anomalies may indicate that the intermediate phase is not a true thermodynamic phase but more likely a non-equilibrium, pinned liquid vortex phase. It is thus more likely to be a glass phase similar to window glass which is an undercooled liquid. This scenario is supported by the observation of supercooling effects in the peak-effect regime [24].

Our scenario is in principle similar to very early work (e.g. Pippard [25] and later the Larkin-Ovchinnikov scenario for collective pinning [26]) where it was proposed that a sudden softening of the vortex shear modulus leads to a better accommodation of the pinning centers by the vortex lattice. However, the driving force for this lattice softening is shown here to be the thermal fluctuations.

Our explanation of the peak effect does not appear to be applicable to the fishtail effect observed in high-temperature superconductors which occurs far from the normal phase boundary. In this case an underlying disorder-induced vortex melting scenario [16] has been proposed, while in our low-$T_c$ superconductor disorder is certainly important to create flux pinning in the peak effect region but clearly not the main driving force. The driving forces here are thermal fluctuations [19] which lead to the observation of a liquid vortex phase [19,27,28]. We would like to mention that a similar 'premelting peak' very close to or even coincidental with thermally induced vortex melting has also been reported in $YBa_2Cu_3O_x$ at fields below the fishtail effect [29,30]. The relaxation behaviour of induced critical currents in the region of this premelting peak shows many features which are known from highly 'fragile' [31] structural glass transitions in undercooled liquids [30]. This all suggests that the presently observed magnetization peak effect in $Nb_3Sn$ (and probably also in many other type-II superconductors) follows the same phenomenology as found in supercooled structural liquids despite the fact that the vortex system is not supercooled in the traditional sense. In the vortex system, the first-order crystallization is prevented not by supercooling, but rather through pinning centers. These can effectively slow down the dynamics of vortex equilibration without significantly disturbing the underlying thermodynamics [19].

## ACKNOWLEDGMENTS

We acknowledge stimulating discussions with T. Giamarchi, M.G. Adesso, R. Flükiger and H. Küpfer. This work was supported by the Swiss National Science Foundation through the National Center of Competence in Research "Materials with Novel Electronic Properties-MaNEP".

**Figure Captions:**

FIG. 1 a) Magnetization vs. magnetic field for various temperatures. The curves are measured for positive and negative $dB_a/dt$, as shown by arrows (see 14 K data). The vertical arrows indicate the upper critical field $H_{c2}$. The small hysteresis loop close to $H_{c2}$ displays the abrupt change from reversible to irreversible behavior which is called the peak effect. The units J/gatT = Am$^2$/gat were chosen so as to be easily comparable with the specific heat measurements. Here 1 gat ('gram-atom') = 1/4 mole. b) Separated reversible ($M_{rev}$, negative signature) and irreversible ($M_{rev}$, positive signature) contribution to the magnetization (see text for details). $H_p$ marks the fields where the critical current density reaches a maximum.

FIG. 2 a) Isothermal magnetocaloric coefficient $M_T$ vs. magnetic field for various temperatures. The curves were measured for positive and negative $dB_a/dt$, as indicated by arrows (see 12 K data). Inset: magnetocaloric coefficient $M_T$ at $T$=12 K for two different field sweep rates ($dB_a/dt$=+0.7 T/min and +1.4 T/min). b) Separated reversible ($M_{T\_rev}$, negative signature) and irreversible ($M_{T\ irr}$, positive signature) contribution to the isothermal magnetocaloric coefficient (see text for details).

FIG. 3 a) Specific heat $C$ measured with an AC technique at constant temperatures during field sweeps. The data has been measured by the same calorimeter under similar conditions to $M_T$. Curves for positive and negative $dB_a/dt$ are observed to merge. $H_l$ marks the field at which a small fluctuation anomaly appears (see text for details). Inset: Superconducting contribution $C^{SC}$ to the specific heat (after subtraction of the constant phonon backgrounds and the Sommerfeld constant) compared with the reversible contribution to the isothermal magnetocaloric coefficient.

FIG. 4 Magnetization $M(H)$, isothermal magnetocaloric coefficient $M_T$ (inverted) and specific heat $C$ (scaled and shifted) at $T$ = 13.5 K plotted simultaneously.

FIG. 5 Inset: $M_T$ with a small (10 G) additional 1 kHz AC field (open triangles) for positive and negative $dB_a/dt$ in comparison with the original data without 'vortex shaking' (closed triangles). $H_{m\ up}$ and $H_{m\ down}$ mark the vortex melting fields dependent on the sign of $dB_a/dt$. For technical reasons the sample was oriented differently with respect to the field in the two experiments [20].

FIG. 6 Phase diagram including $H_P$ (the maximum of the peak effect as obtained from the magnetization and magnetocaloric coefficient), the lower and upper onset of the peak effect from the magnetization and magnetocaloric coefficient, $H_l$ (maximum of the fluctuation lambda anomaly from specific heat $C(H)$), $H_N$ (fields above which the superconducting contributions to the specific heat vanish) and $H_m$ (vortex melting transition from specific heat $C(T)$) measured conventionally during a temperature sweep [19] and magnetocaloric coefficient ($H_{m\ up}$ and $H_{m\ down}$ for positive and negative $dB_a/dt$ values).

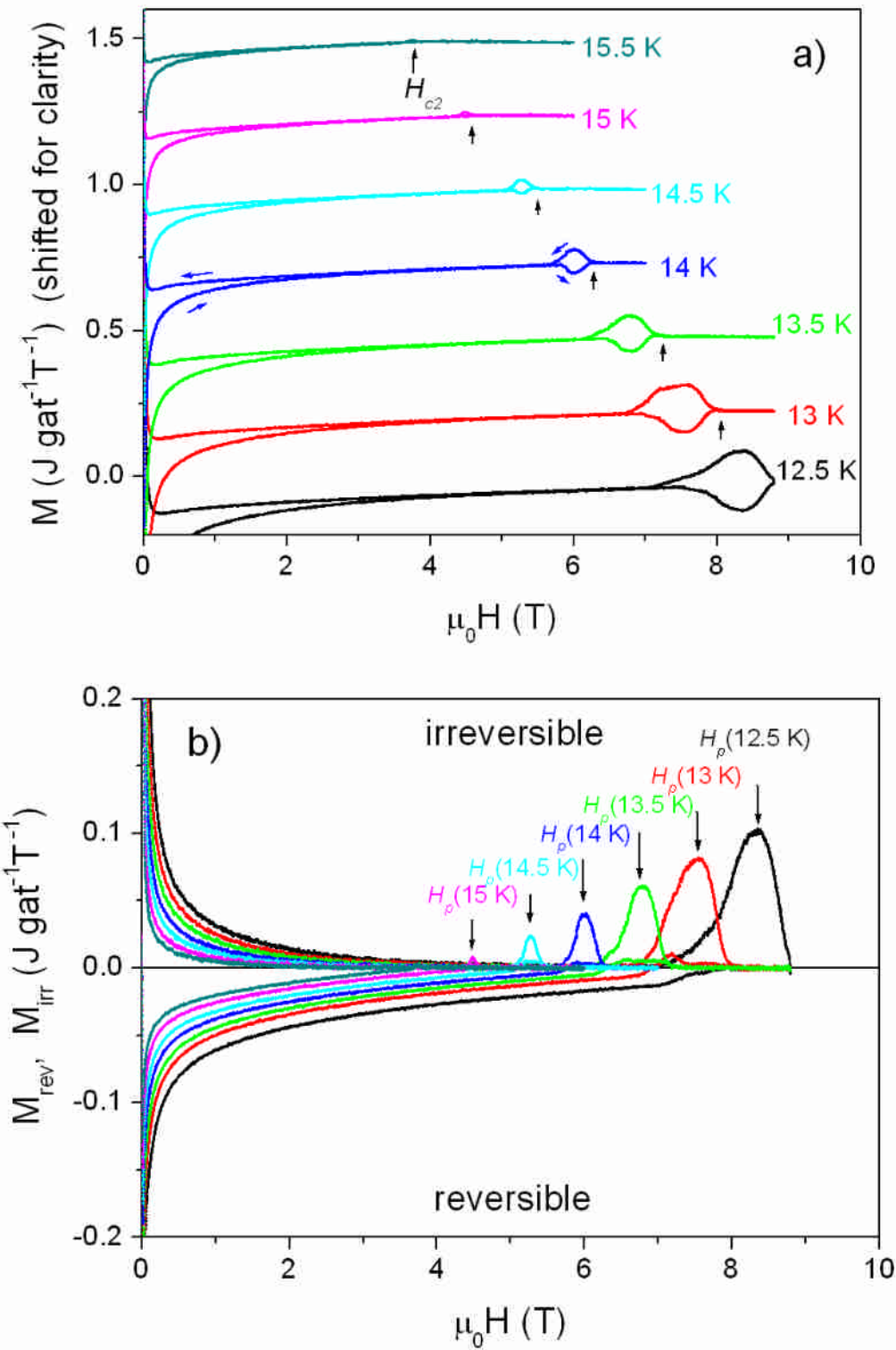

**FIG 1a**
**FIG 1b**

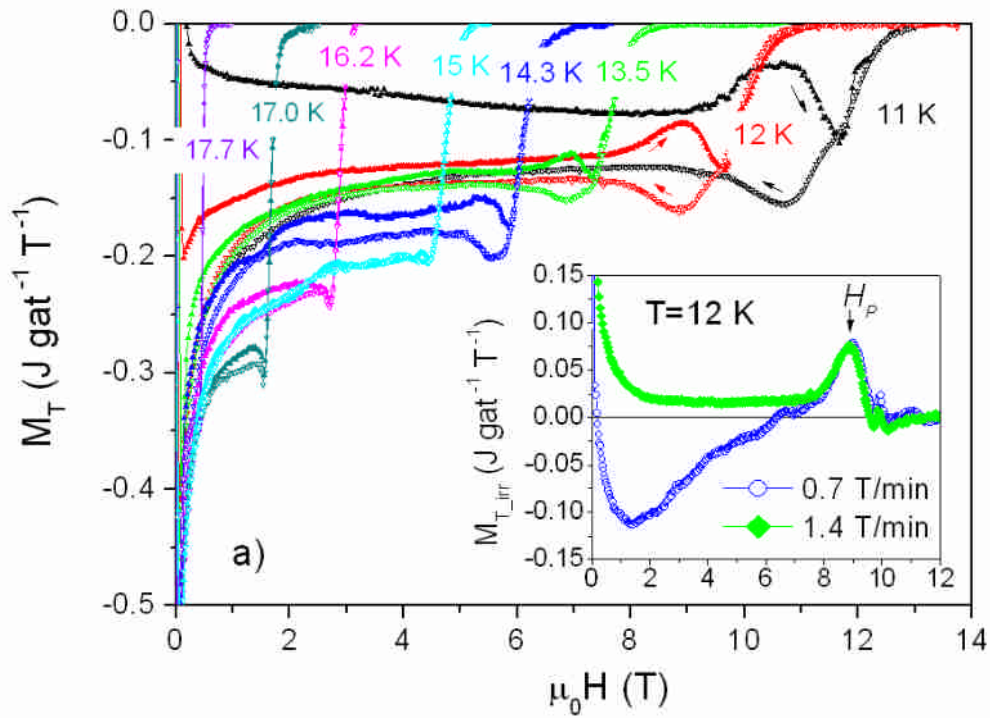
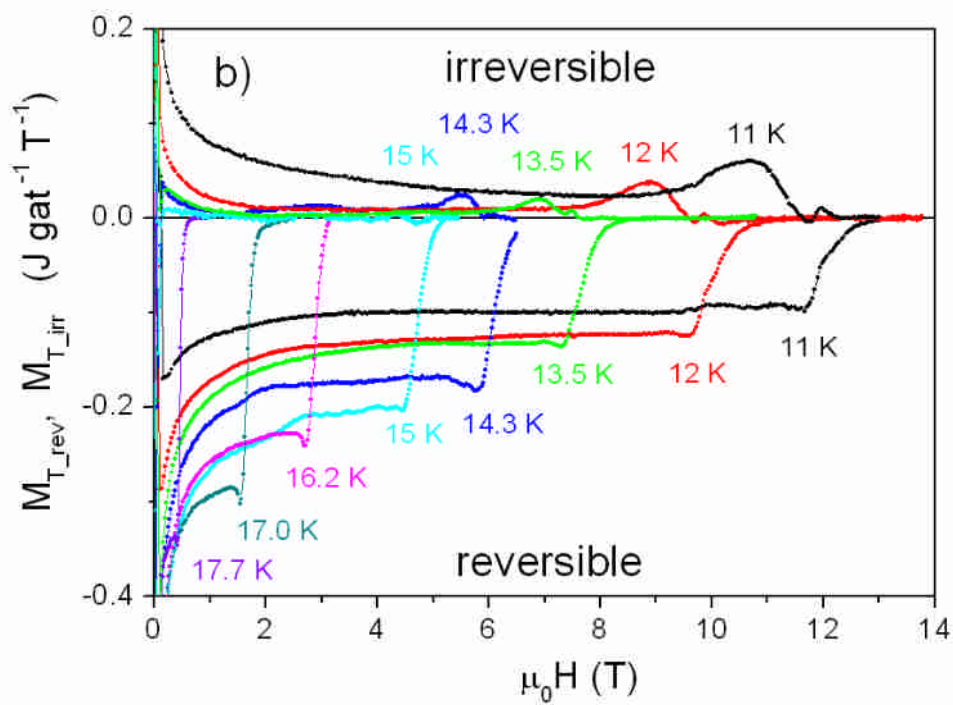

**FIG 2a**
**FIG 2b**

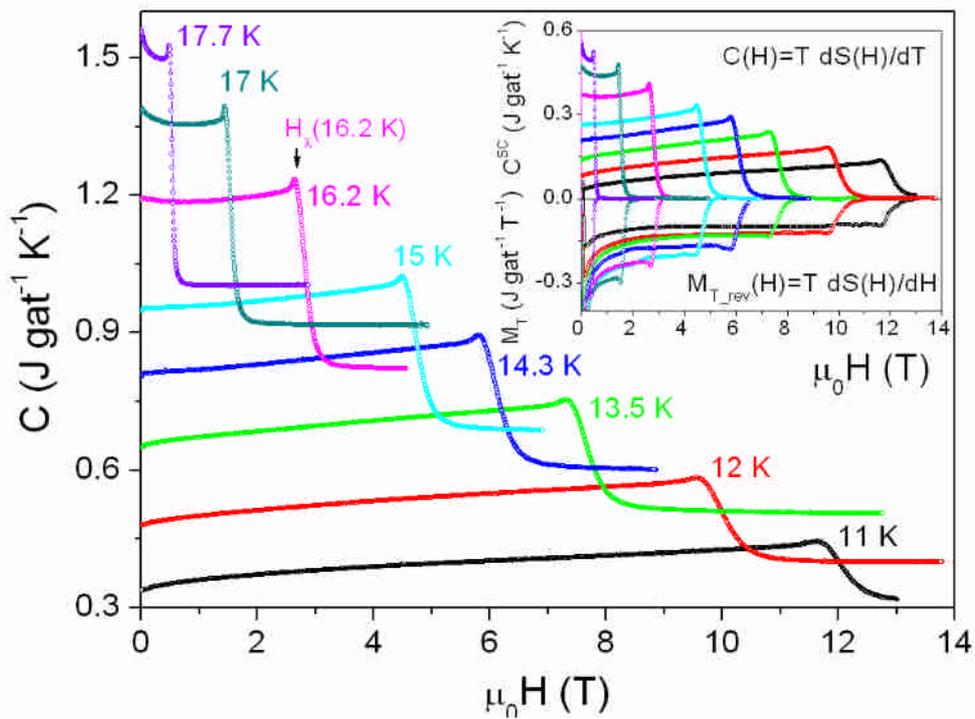

**FIG 3**

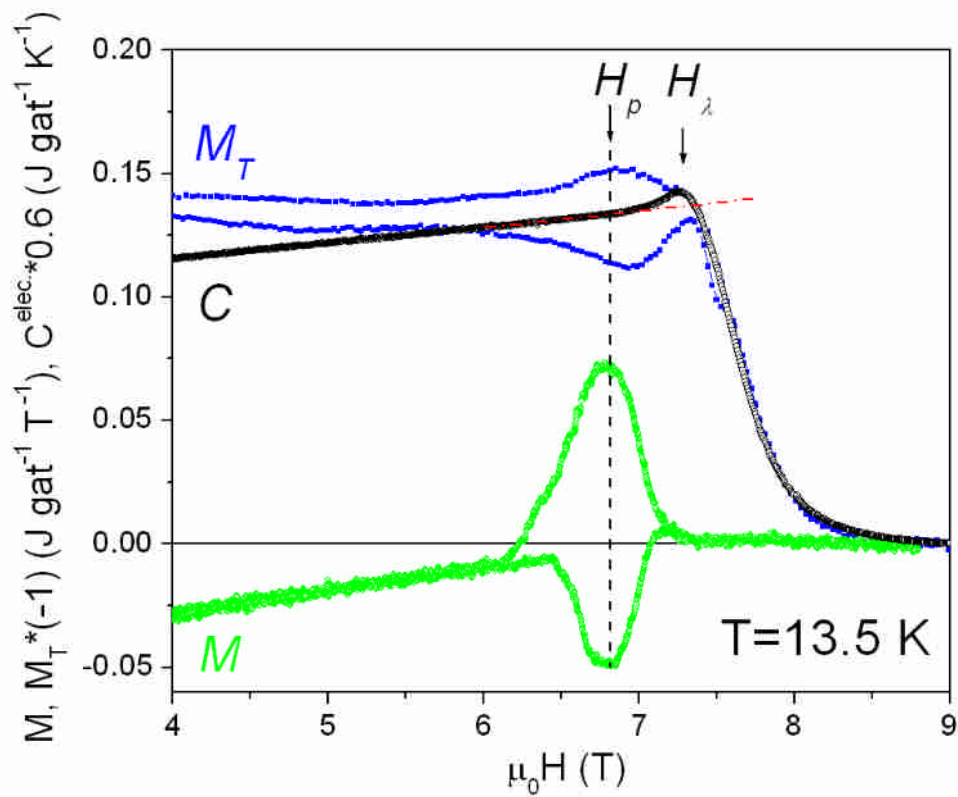

**FIG 4**

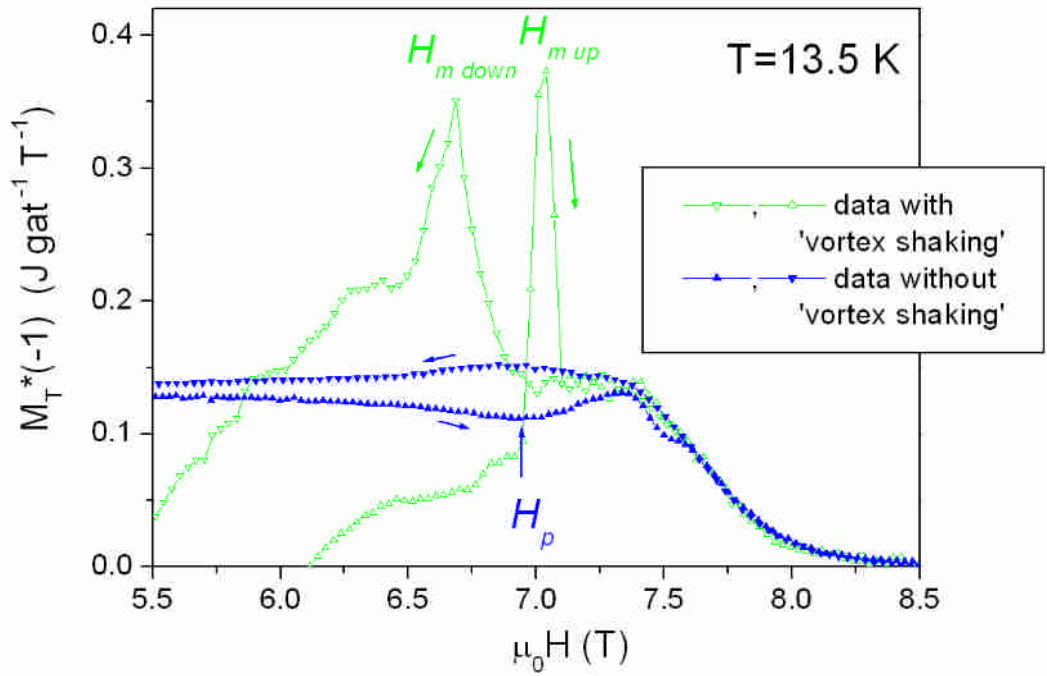

**FIG 5**

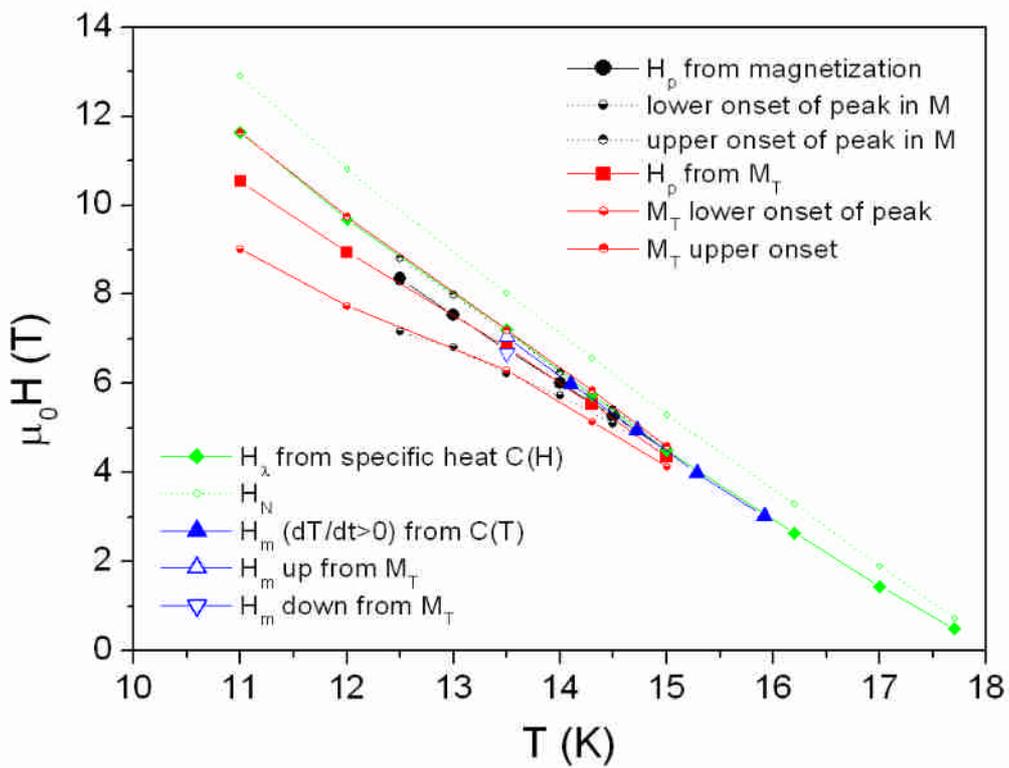

**FIG 6**